\documentclass{ws-ijmpcs}
\usepackage{epsfig,graphicx}
\def\trimmarks{}

\begin{document}

\markboth{I. Vitev}
{QCD evolution techniques for heavy ion physics}

%
\catchline{}{}{}{}{}
%

\title{Hard probes in heavy ion collisions: current status \\ and prospects for application of QCD evolution techniques}

\author{Ivan Vitev}

\address{Theoretical Division, Los Alamos National Laboratory, Mail Stop B283 \\
Los Alamos, NM 87545,
U.S.A. \\
ivitev@lanl.gov}

\maketitle

\begin{history}
\received{Day Month Year}
\revised{Day Month Year}
\published{Day Month Year}
\end{history}

\begin{abstract}
In the past decade the observation of cross section modification for leading hadrons, heavy 
flavor and two particle correlations in heavy ion collisions has provided important insights into the dynamics
of parton propagation in dense strongly-interacting matter. The development of the theory of 
reconstructed jets and related experimental measurements have further shed light on the characteristics 
of in-medium parton showers. So far, experimental results from ultra-relativistic nuclear collisions at RHIC
and LHC have been analyzed in the framework of parton energy loss, where the precision of the theoretical 
predictions cannot be systematically  improved. Only recently have higher order calculations and 
applications of resummation  and evolution to heavy ion collisions begun to emerge. Several examples of 
such advances  are discussed in these proceedings.

\end{abstract}

\ccode{PACS numbers:12.38.-t, 12.38.Bx,  12.38.Cy, 24.85.+p}

\section{Introduction}	

In this brief overview, emergent approaches that aim at increasing the theoretical precision in the evaluation of hard probes observables in heavy ion collisions are discussed. It was realized more than twenty years ago that in the ambiance of strongly-interacting matter the cross sections for high transverse momentum hadron production will be attenuated. This phenomenon, dubbed "jet quenching", has attracted tremendous attention from the theoretical and experimental communities~\cite{Gyulassy:2003mc} at the Relativistic Heavy Ion Collider (RHIC) and the Large Hadron Collider (LHC) and has subsequently been extended to heavy flavor production, particle correlations and, most importantly, reconstructed jet observables. Within the framework of the traditional parton energy loss approach, however, the accuracy of such theoretical analyses cannot be systematically improved. Recent advances in the theory of in-medium parton shower formation allow us to overcome this limitation and to unify our understanding of energetic particle and jet production in hadronic and nuclear collisions.       
Next-to-leading order calculations and application of evolution techniques for deep inelastic scattering  and heavy ion collisions at 
RHIC and the LHC are described in more detail in \cite{xing,Kang:2014ela,Ovanesyan:2014mfa,Kang:2014xsa,chien,Chien:2014nsa}.
A bier overview of soft-collinear effective theory (SCET) and discussion of the connection between perturbative QCD (pQCD) and SCET resummation techniques can be found in \cite{Lee:2014xit,Almeida:2014uva}. For applications to heavy ion collisions, SCET has been extended to include Glauber modes~\cite{Idilbi:2008vm,Ovanesyan:2011xy}. The progress toward better understanding of hard probes in heavy ion collisions, with emphasis on future directions,  is concisely reviewed below.

\section{Parton propagation in matter and energy loss}

Given the significance of understanding the production of large transverse momentum particles and jets
in the ambiance of the QGP, several theoretical formalisms were developed to evaluate the energy loss 
of quarks and gluons as they propagate through the medium~\cite{Baier:1998kq,Gyulassy:2000er,Wang:2001ifa,Arnold:2002ja}
As an example, we present below the Guuylassy-Levai-Vitev (GLV) reaction operator approach where the fully differential
medium induced gluon bremsstrahlung intensity is obtained as a series over the multiple scattering correlations in the
medium.  For gluons with momentum $k=[xp^+, {\bf k}^2 / xp^+,{\bf k}]$  resulting from the sequential interactions 
of a fast parton with momentum $p=[p^+, 0,{\bf 0}]$  can be written as 
\begin{eqnarray}
 x\frac{dN_g}{dx\, d^2 {\bf k}}  &=&
\sum\limits_{n=1}^\infty  x\frac{dN_g^{(n)}}{dx\, d^2 {\bf k}}   
 = \sum\limits_{n=1}^{\infty}  \frac{C_R \alpha_s}{\pi^2} 
 \; \prod_{i=1}^n \;\int_0^{L-\sum_{a=1}^{i-1} \Delta z_a } 
 \frac{d \Delta z_i }{\lambda_g(i)}  \, 
  \nonumber \\[1.ex]   
 &\;&   \int   d^2{\bf q}_{i} \, 
\left[  \sigma_{el}^{-1}(i)\frac{d \sigma_{el}(i)}{d^2 {\bf q}_i}
  - \delta^2({\bf q}_{i}) \right]  \,   \left( -2\,{\bf C}_{(1, \cdots ,n)} \cdot 
\sum_{m=1}^n {\bf B}_{(m+1, \cdots ,n)(m, \cdots, n)}   \right. 
\nonumber \\[1.ex]   
 &\;&  
\left. \times \left[ \cos \left (
\, \sum_{k=2}^m \omega_{(k,\cdots,n)} \Delta z_k \right)
-   \cos \left (\, \sum_{k=1}^m \omega_{(k,\cdots,n)} \Delta z_k \right)
\right]\; \right) \;, \quad \qquad  
\label{difdistro} 
\end{eqnarray}
where $\sum_2^1 \equiv 0$ is understood. In the small angle 
eikonal limit 
$x=k^+/p^+ \approx \omega/E$. In Eq.~(\ref{difdistro})  
the color current propagators are denoted by
\begin{eqnarray}
{\bf C}_{(m, \cdots ,n)} &=&  \frac{1}{2} \nabla_{{\bf k}} 
\ln \, ({\bf k} - {\bf q}_m - \cdots  - {\bf q}_n )^2 \nonumber \\  
{\bf B}_{(m+1, \cdots ,n)(m, \cdots, n)} &=&  
{\bf C}_{(m+1, \cdots ,n)} - {\bf C}_{(m, \cdots ,n)} \;\;. 
\label{props}
\end{eqnarray}
The momentum transfers ${\bf q}_i$ are distributed according to 
a normalized elastic differential cross section,  
\begin{equation}
\sigma_{el}(i)^{-1}\frac{d \sigma_{el}(i)}{d^2 {\bf q}_i}  
= \frac{\mu^2(i)}{\pi({\bf q}_i^2+\mu^2(i))^2} \; ,
\label{GWmodel}
\end{equation}
which models scattering  by  soft partons with a thermally 
generated Debye screening mass $\mu(i)$. 
 For gluon dominated bulk soft matter  $\sigma_{el}(i)  \approx  \frac{9}{2} \pi \alpha_s^2/\mu^2(i)$ 
and  $\lambda_g(i)=1/\sigma_{el}(i)\rho(i)$.   The characteristic path length dependence of the non-Abelian energy 
loss in Eq.~(\ref{difdistro}) comes from the interference phases and is differentially controlled
by the inverse formation times,    
\begin{equation}
\omega_{(m,\cdots,n)}  = 
\frac{({\bf k} - {\bf q}_m - \cdots  - {\bf q}_n )^2}{2 x E}  \;, 
\label{ftimes}
\end{equation}
and the separations of the subsequent scattering centers  
$ \Delta z_k = z_k - z_{k-1} $. It is the non-Abelian 
analogue of the Landau-Pomeranchuk-Migdal destructive interference effect in QED, which, for a static plasma, up to 
kinematic corrections yields the following parametric behavior with respect to the energy and path length of the parton: 
$\Delta E \sim   L^2 \ln E$.

\begin{figure}[!t] 
\begin{center}
\hspace*{-0.5cm }\epsfig{file=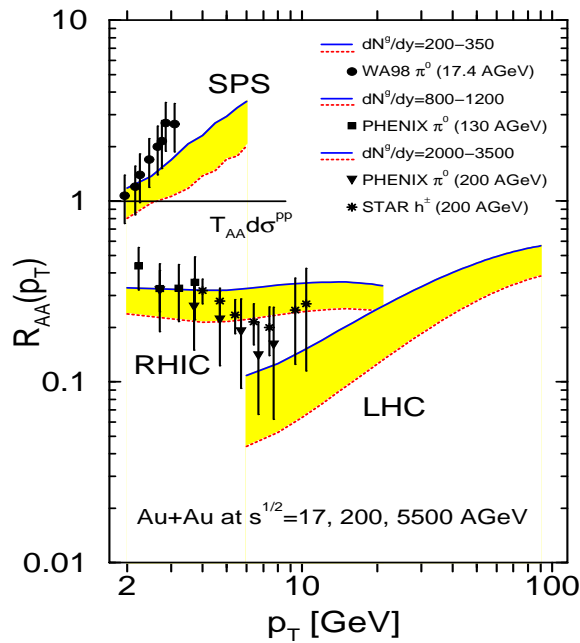,
height=3.in,width=3.4in,clip=5,angle=-90}  
\vspace*{0.cm} 
\caption{The suppression/enhancement ratio $R_{AA}(p_T)$  
         ($A=B=Au$)   for neutral 
         pions at $\sqrt{s}_{NN}=17$, $200$, $5500$~GeV.   Solid (dashed) 
         lines correspond to the smaller (larger)  effective initial  
         gluon rapidity densities at given $\sqrt{s}$ that drive parton 
         energy loss. Data on
         $\pi^0$ production in central $Pb+Pb$ at $\sqrt{s}_{NN}=17.4$~GeV 
         from WA98 and in central $Au+Au$ at
         $\sqrt{s}_{NN}=130$~GeV, as well as  
         {\em preliminary} data at 
         $200$~GeV  from PHENIX and 
         $h^\pm$ central/peripheral data from 
         STAR are shown. The sum of 
         estimated statistical and systematic errors are indicated. }  
\label{sup_sys} 
\end{center}
\end{figure} 

\subsection{The energy loss approach to jet quenching}

Should an energy loss approach be adopted,  it is  important to realize that  the soft gluon emission limit 
must be consistently implemented. If the  fractional energy loss  becomes significant, it is carried away through 
multiple gluon bremsstrahlung. In the independent Poisson gluon emission  limit, we  can construct the probability density  
 $P_c(\epsilon)$ of this  fractional energy loss   $\epsilon  = \sum_{i } \omega_i /E \approx   \sum_{i } k^+_i/p^+$,
 such that: 
 \begin{eqnarray} 
&& \int_0^1 d \epsilon \; P( \epsilon ) = 1\;, \quad 
 \int_0^1 d \epsilon \; \epsilon \, P( \epsilon ) = 
\left\langle  \frac{ \Delta E }{ E } \right\rangle \;. \quad 
\label{meane} 
\end{eqnarray}
A more detailed discussion is given in~\cite{Gyulassy:2003mc}.  If a parton loses this energy fraction $\epsilon$ 
during its propagation  in the QGP to escape with momentum $p^{\rm quench }_{T_c}$, immediately 
after the hard collision  $p_{T_c} = p^{\rm quench}_{T_c} /  ( 1 - \epsilon )$. Noting the additional 
Jacobian $|d p_{T_c}^{\rm quench} 
/ d p_{T_c}| = ( 1- \epsilon) $, the kinematic modification to the FFs due to energy loss is:
\begin{eqnarray}
&&D_{c}^{\rm quench}(z) =     \int_0^{1-z} \, d \epsilon\,   \frac{ P_c(\epsilon)  }{(1-\epsilon)}
D_{c}\left(\frac{z}{1-\epsilon}\right)  \, , \quad
\label{quenchD}
\end{eqnarray}
and can be directly implemented in the evaluation of inclusive and tagged hadron production. 
Comparison of theoretical predictions for the  nuclear modification factor for inclusive light hadron production, defined as
\begin{eqnarray} 
&&R_{AA} (p_T)  = 
\frac{ d \sigma^h_{AA}/dyd^2 p_T }
{   \langle N_{ \rm coll}\rangle    d \sigma^h _{pp}  / dyd^2 p_T } \, , \label{raa-analyt} 
\label{raadef}
\end{eqnarray}
to experimental data is shown in Fig.~\ref{sup_sys}. In Eq.~(\ref{raadef})  ${\langle N_{ \rm coll}\rangle}$ is 
the average number of binary nucleon-nucleon collisions.  

A wide variety of models for parton energy loss were employed to fit RHIC "jet quenching" measurements and theoretical 
predictions for the Pb+Pb run at the LHC were summarized in~\cite{Armesto:2009ug}. The majority of these calculations 
failed to describe  the LHC inclusive particle suppression data~\cite{CMS:2012aa}, in particular its $p_T$ dependence, which
Fig.~\ref{sup_sys} represents well.   Although it was 
subsequently shown that many of the approaches can fit the data through model improvements and parameter adjustment, it
is now clear that better understanding of the characteristics of the in-medium and improved theoretical precision of the 
calculations is necessary.

\subsection{The theory of reconstructed jets in heavy ion collisions}

Suppression of the reconstructed jet cross sections and modification of the jet substructure observables 
in heavy ion collisions are sensitive signatures of the QGP-modified parton dynamics. In particular, jet measurements
in heavy ion collisions provide unique access to the longitudinal and transverse structure of the medium induced
parton shower that is not possible through leading particle or particle correlations analyses. For a jet reconstruction
 parameter $R$ one can evaluate from Eq.~(\ref{difdistro})
the part of  the fractional energy loss $\epsilon_i$ that is simply redistributed inside the jet for a parent parton "i" 
\begin{eqnarray}
 f(R_i,p_{T\, i }^{\min})_{q,g}= \frac{\int_0^{R_i} dr \int_{p_{T\,i}^{\min}}^{E_{T\,i}} d\omega \,
\frac{dI^{\rm rad}_{q,g}(i)}{d\omega  dr }} {\int_0^{{R\, i}^{\infty}} dr \int_{0}^{E_{T\,i}} d\omega \,
\frac{dI^{\rm rad}_{q,g}(i)}{d \omega dr} } \; .
\label{fraction}
\end{eqnarray}
Thus,  $f(R_i,p_{T\, i }^{\min})_{q,g}$ plays a critical role in defining the contribution
(or lack thereof  if $f(R_i,p_{T\, i }^{\min})_{q,g} \rightarrow 0 $)
of  the  medium-induced  bremsstrahlung  to  the  jet. Note that in Eq.~(\ref{fraction}) $p_{T\, i}^{\min}$ is a parameter 
that can be used to simulate processes that can alter the energy
of the jet beyond medium-induced bremsstrahlung, for example collisional energy loss.

The medium-modified jet cross section for jets per binary scattering is evaluated as
\begin{eqnarray}
&&\frac{1}{\langle  N_{\rm bin}  \rangle}
 \frac{d\sigma^{AA}(R)}{dy_1dE_{T\,1}} =  \sum_{q,g}
\int_{\epsilon_1=0}^1 d\epsilon_1 \;   \frac{P_{q,g}(\epsilon_1,E_{T\,1}) }{ \left(1 - [1-f(R_1,p_{T\,1}^{\min})_{q,g}]  \epsilon_1\right)}
 \frac{d\sigma^{\rm CNM,NLO}_{q,g}(E_{T\,1}^\prime)} {dy_1 dE^{\, \prime}_{T\,1}} \; . \qquad
\label{incl}
\end{eqnarray}
In Eq.~(\ref{incl}) the phase space Jacobian reads
$$|J_i(\epsilon_i)| = 1/\left(1 - [1-f(R_i,p_{T\,i}^{\min})_{q,g}]  \epsilon_i\right)\, , $$
and  $E_{T\,i}^\prime = |J_i(\epsilon_i)|E_{T\,i}$ in the argument of $\sigma^{\rm CNM,NLO}_{q,g}$.

\begin{figure}[!t]
\vskip0.04\linewidth
\centerline{
\includegraphics[width = 0.45\linewidth]{LHCcouplingCH.eps} \includegraphics[width = 0.55\linewidth]{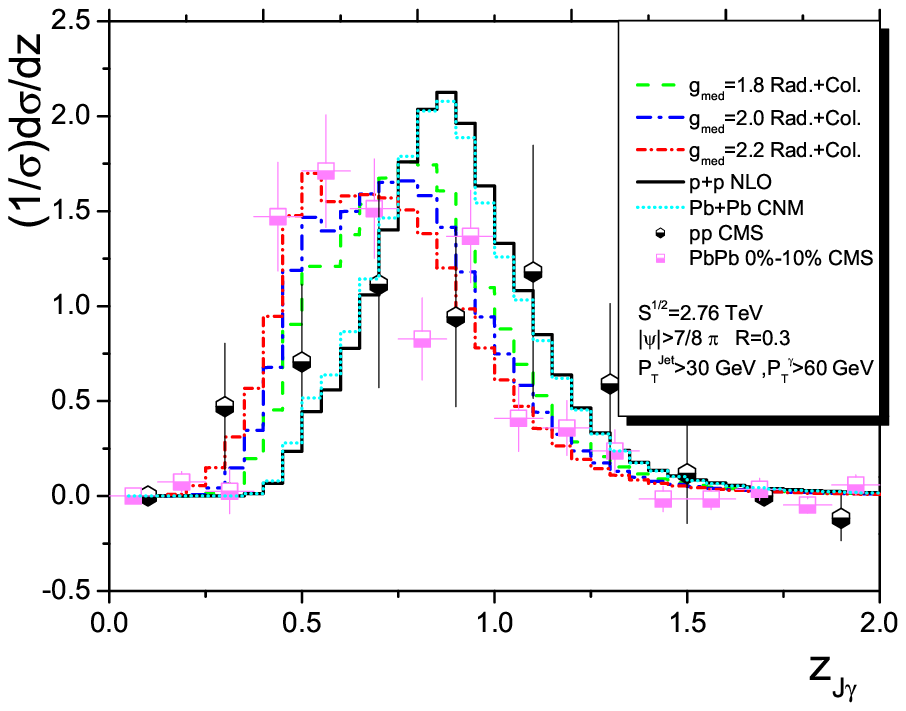} 
}
\caption{ Left panel: the $E_T$ dependence of the nuclear modification factor for different
jet cone sizes  $R=0,2, \, 0.6$ is calculated in central Pb+Pb collisions at the
LHC $\sqrt{s_{NN}}=2.76$~TeV. Bands represent the variation in the coupling strength between
the jet and the medium. The relative contribution of cold nuclear matter effects
to $R_{AA}$ is illustrated for $R=0.2, \, 0.4, \, 0.6, \, 0.8$. Right panel:  the momentum imbalance of photon tagged jets in 
p+p and A+A collisions is compared to CMS experimental data in p+p and Pb+Pb reactions.}
 \label{RAAlhccoupling}
\end{figure}

One of the most important theoretical predictions related to jet production in heavy ion collisions is the  radius dependence 
of jet suppression~\cite{He:2011pd}. It is illustrated in the left panel of Fig.~\ref{RAAlhccoupling} and arises from the 
broader angular distribution of the QGP-induced parton shower in comparison to the parton shower in the absence of 
the strongly-interacting medium. In the limit of small jet radii $R \rightarrow 0$, it 
approaches the level of suppression observed in inclusive particle production. This expectation is now confirmed by the ATLAS 
experimental data~\cite{Aad:2012vca}.  Further constraints on the modification of the parton shower can be placed through electroweak boson tagging~\cite{Neufeld:2010fj} and the change in the shape and magnitude of correlation distribution, such as the momentum imbalance
$z_J = p_{T\, {\rm jet}}/p_{T\, \gamma,Z^0}$. One such example is shown in the right panel of Fig.~\ref{RAAlhccoupling} and
additional theoretical predictions for these channels have also been confirmed by LHC experimental data.

\section{Resummation for hard probes in heavy ion collisions}

The next step in improving the theory if leading particles and jets in heavy ion reactions is to employ resummation/evolution 
techniques where the theoretical control on the accuracy of the evaluation is much improved. An important step forward was the
derivation of all medium-induced splitting functions beyond the soft-gluon (energy loss) approximation \cite{Ovanesyan:2011kn}.

\subsection{Evolution approach to leading particle production in A+A}

 A critical step in improving the jet quenching phenomenology is to understand the implication of the finite-$x$ corrections. Their implementation
requires new theoretical methods, since in the large momentum fraction  limit the leading parton can change flavor and the splitting process cannot be interpreted as  energy loss. A natural language to capture this physics is that of  the well-known  Dokshitzer-Gribov-Lipatov-Altarelli-Parisi  (DGLAP)  evolution equations, which provide up to modified leading logarithmic accuracy.  As a first application of the full medium-induced splitting kernels~\cite{Ovanesyan:2014mfa,Ovanesyan:2011kn}, we revisit the evaluation of the 
nuclear modification factor $R_{AA}$ for inclusive hadron production at high transverse momentum $p_T$ 
(and rapidity $y$)~\cite{Kang:2014xsa}. 

\begin{figure}[!t]
\begin{center}
\includegraphics[width=8cm]{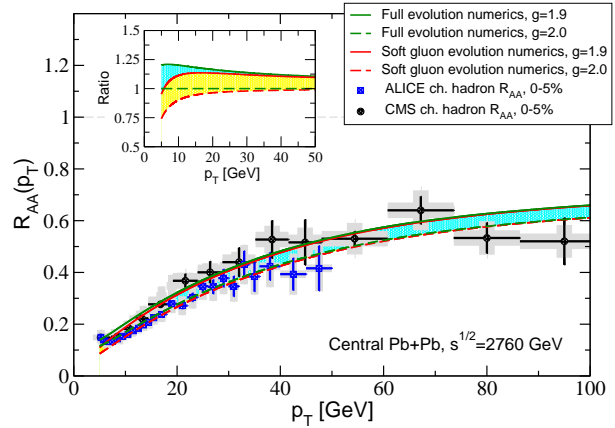}
\vspace{5mm}
\caption{ Comparison between $R_{AA}$  obtained with in-medium numerically evolved fragmentation functions using the full splitting kernels 
 (cyan band) and their soft gluon limit (yellow band) to ALICE and CMS data. The upper and lower edges of the bands correspond to $g=1.9$ and $g=2.0$, respectively.
}\label{fig:evolutionRAA2}
\vspace{0mm}
\end{center}
\end{figure}

Given the success of the traditional jet quenching phenomenology, its is important to establish the connection between the evolution
and energy loss approaches \cite{Ovanesyan:2014mfa,Kang:2014xsa}. This can be achieved {\em only} in the soft gluon 
bremsstrahlung  limit,  where the two diagonal splitting functions $P_{q\rightarrow qg}(z')$ and $P_{g\rightarrow gg}(z')$ survive. 
  branching is given by a plus function. The DGLAP evolution equations decouple and because the fragmentation functions $D(z/z')$ are typically steeply falling with increasing $p_T^{\rm hadron}/ p_T^{\rm parton}$, the main 
contribution   comes predominantly from $z'\approx 1$. We expand the integrand in this limit,  the analytical solution 
\begin{equation}
D^{\text{med}}(z,Q)\approx e^{-(n(z)-1)\left\langle\frac{\Delta E}{E}\right\rangle_z-\left\langle N_g\right\rangle_z}D^{\rm vac}(z,Q)\,, \label{eq:masterconnect}
\end{equation}
and shows explicitly that the vacuum evolution and the medium-induced evolution factorize.
We have used the following definitions in the above formula:
\begin{equation}
 \left\langle\frac{\Delta E}{E}\right\rangle_z=\int_0^{1-z}d x \,x\,\frac{d N}{d x}(x) \xrightarrow{z \to 0}\left\langle\frac{\Delta E}{E}\right\rangle\, ,  \; \left\langle N_g\right\rangle_z=\int_{1-z}^1d x\,\frac{d N}{d x}(x)\xrightarrow{z\to1}\left\langle N_g\right\rangle\,, \quad
\end{equation}
where $xdN/dx$ is the medium-induced gluon intensity distribution~\cite{Gyulassy:2000er} and the connection to the traditional energy loss approach is evident.

Conversely, solving the DGLAP evolution equations with the full medium induced splitting functions allows us to unify the treatment of vacuum and medium-induced parton showers. It gives a very good description of the RHIC and LHC experimental data, as illustrated in Fig.~\ref{fig:evolutionRAA2}. We find that the
coupling between the jet and the medium can be constrained with better than 10\% accuracy when the uncertainties that arise from 
the choice of method and the fit to the data are combined. The investigation of the medium-evolved fragmentation functions further reveals that significant improvement in the reliability of the theoretical predictions by using the  approach described here can be obtained for more exclusive observables.

\subsection{Jet shapes in hadronic collisions}

The jet shape is a classic jet substructure observable that probes the average transverse energy profile inside a reconstructed jet. 
Given a jet with an axis $\hat n$, its integral and differential jet shapes, $\Psi_J(r)$ and $\psi_J(r)$ ,  are defined as follows,
\begin{equation}
    \Psi_J(r)=\frac{\sum_{i,~d_{i\hat n}<r} E^i_T}{\sum_{i,~d_{i\hat n}<R} E^i_T}\;,  \quad
 \psi_J(r)=\frac{ d\Psi_J(r)}{dr}  \;.
\end{equation}
The studies of jet shapes in proton-proton collisions have served as precision tests of perturbative Quantum Chromodynamics (QCD). They have also recently become the baseline for studying the in-medium modification of parton showers in ultra-relativistic nucleus-nucleus collisions. The jet shape is a function of two angular parameters $R$ and $r$, which can be at hierarchical scales. Its calculation suffers from large logarithms of the ratio between the two scales, and these phase space logarithms can be conveniently resummed in the framework of soft-collinear effective theory (SCET). Early evaluation of jet shapes  in heavy ion collisions~\cite{Vitev:2008rz} has predicted all qualitative features of the modification of this observable relative to p+p collisions, measured only recently in Pb+Pb reactions at the LHC~\cite{Chatrchyan:2013kwa}.  At the same, time the need for quantitative improvement in theory is significant. Part of this improvement is a better evaluation of the  $\psi_J(r)$ p+p baseline.

\begin{figure}[t]
    \begin{center}
    \includegraphics[scale=0.28, trim = 0mm 0mm 0mm 0mm , clip=true]{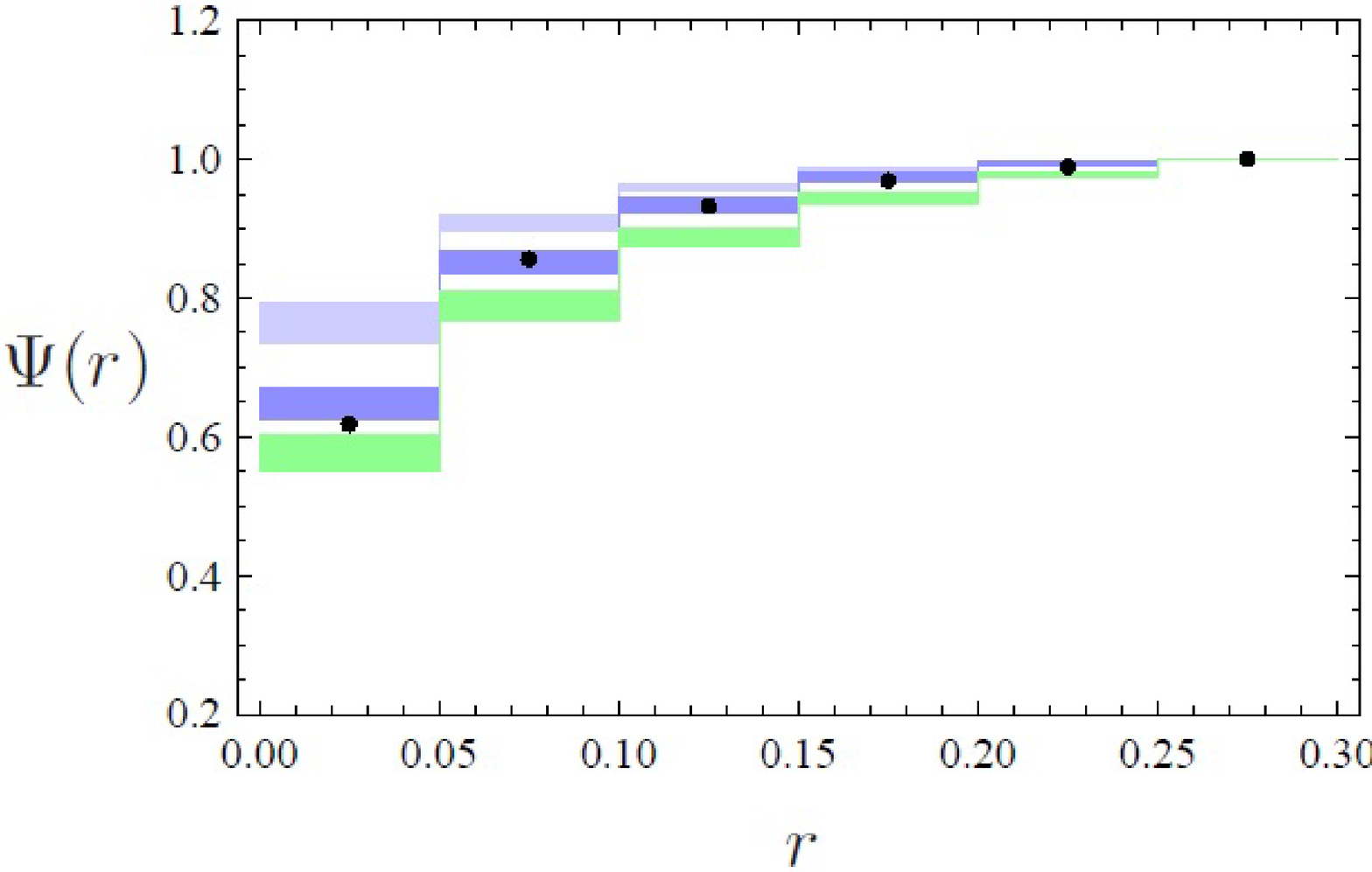}
    \includegraphics[scale=0.28, trim = 0mm 0mm 0mm 0mm , clip=true]{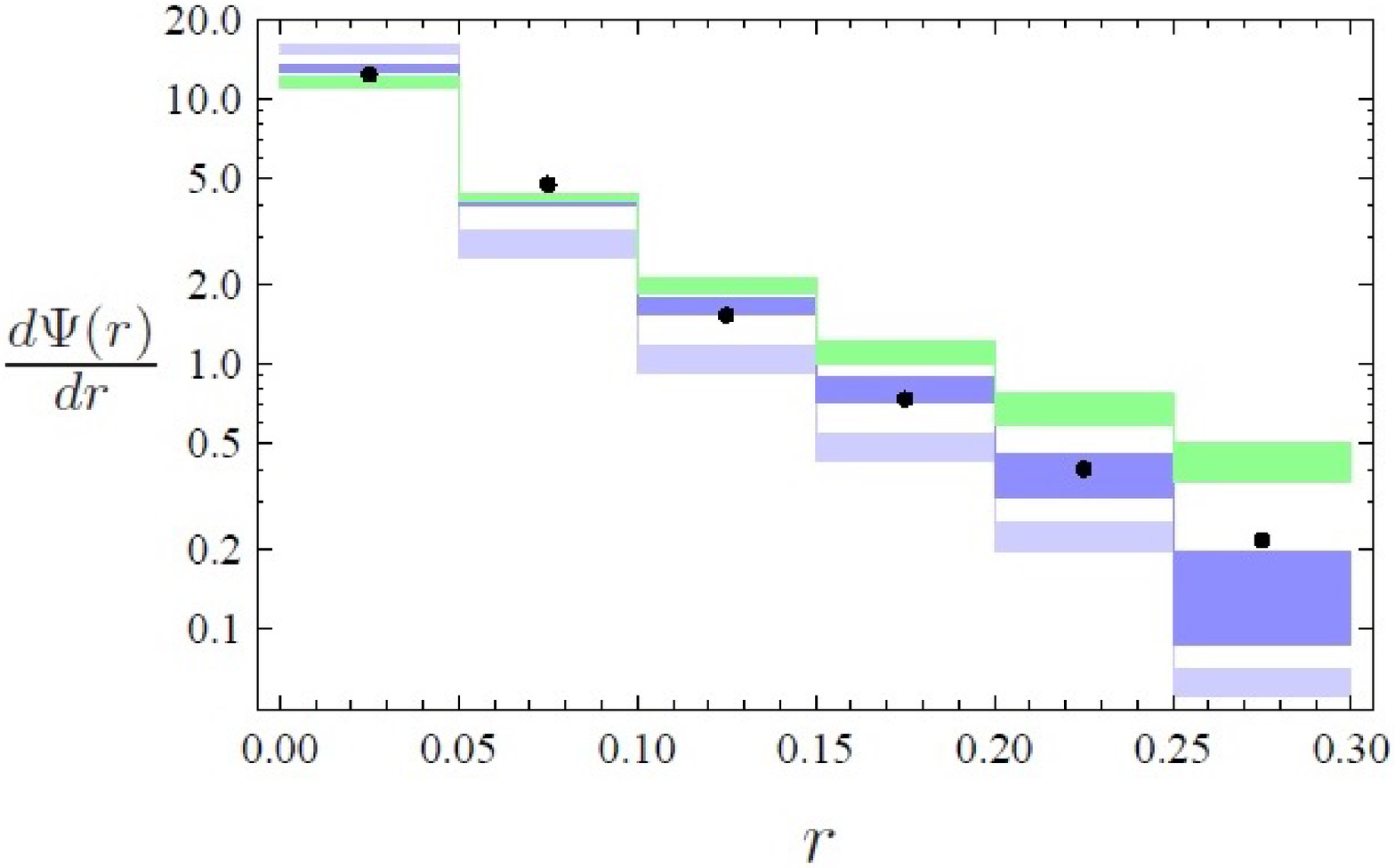}
    \caption{\label{fig:result} The integral (left) and differential (right) jet shapes in proton-proton collisions with center of mass energy at $\sqrt{s_{\rm NN}}=2.76$ TeV are plotted as a function of $r$, which is the angle from the jet axis. Jets are reconstructed using the anti-$\rm k_T$ algorithm with $R=0.3$. The cuts on the transverse momenta and rapidity of the jets ($p_T^{\rm jet}>100$ GeV and $0.3<|y^{\rm jet}|<2$) are imposed. The dots are the CMS data with negligible experimental uncertainties. The shaded blue boxes are the LO (light) and NLL (dark) results for anti-$\rm k_T$ jets, with the theoretical uncertainties estimated by varying the jet scales between $\frac{1}{2}\mu_{j_R}<\mu<2\mu_{j_R}$. The NLL results agree with the data much better than the LO results. The shaded green boxes are the NLL results for cone jets, plotted as an illustration of the algorithm dependence in jet shapes. }
    \end{center}
\end{figure}

The evaluation of jet shapes in the framework of SCET, up to  power corrections from initial state radiation,  was presented in \cite{chien,Chien:2014nsa}. The SCET calculations of the differential jet shapes $\Psi_\omega(r)$ includes the one-loop jet energy functions, the two-loop cusp anomalous dimensions ($\Gamma_0$ and $\Gamma_1$) and the one-loop anomalous dimensions ($\gamma_0^{q,g}$) of the jet energy functions, as well as the two-loop running of the strong coupling constant with $\alpha_s(m_Z)=0.1172$. This will give us the precision formally at next-to-leading logarithmic order (NLL), including terms down to $\alpha_s^n\ln^{2n-1}\frac{r}{R}$.

Figure \ref{fig:result} shows the comparison between the LO and NLL calculations and the CMS measurement of the integral and differential jet shapes in proton-proton collisions at $\sqrt{s_{\rm NN}}=2.76$ TeV. The shaded boxes are the theoretical uncertainties we estimate by varying the jet scales between $\frac{1}{2}\mu_{j_R}<\mu<2\mu_{j_R}$.  The LO calculation, due to its divergent nature, certainly can not describe the data and resummation becomes necessary. The results for cone jets are also shown to illustrate the algorithm dependence in jet shapes. 
 
\section{Conclusions}

These proceedings presented a brief overview of the evolution of medium-induced parton energy loss "jet quenching" applications  form leading particles to jets. I discussed emergent approaches, based or resummation/evolution and modern effective theories of QCD, that aim at increasing the theoretical precision in the evaluation of hard probes observables in heavy ion collisions. These theoretical advances pave the way toward a unified treatment of vacuum and in-medium parton showers and a common approach to understanding jet production in article and high-energy nuclear physics.

\section*{Acknowledgments}

I. Vitev is supported by the US Department of Energy, Office of Science, Office of Nuclear Physics and Office
of Fusion Energy Sciences and in part by the LDRD program at LANL.

\end{document}